\documentclass[useAMS,usegraphicx,usenatbib]{mn2e}
\usepackage{multirow}

\title[Spatial dependence of the SFH in the Fornax dSph]{Spatial dependence of the Star Formation History in the Central Regions of the Fornax Dwarf Spheroidal Galaxy}
\author[Andr\'es del Pino et al.]{Andr\'es del Pino$^{1}$\thanks{E-mail: \mbox{adpm@iac.es} (ADPM); \mbox{shidalgo@iac.es} (SH); \mbox{aaj@iac.es} (AA); \mbox{carme@iac.es} (CG); \mbox{rcarrera@iac.es} (RC); \mbox{monelli@iac.es} (MM); \mbox{buonanno@oa-teramo.inaf.it} (RB); \mbox{gmarconi@eso.org} (GM)}, Sebastian L. Hidalgo$^{1,2}$\footnotemark[1], Antonio Aparicio$^{1,2}$\footnotemark[1],\newauthor
 Carme Gallart$^{1,2}$\footnotemark[1],Ricardo Carrera$^{1,2}$\footnotemark[1], Matteo Monelli$^{1,2}$\footnotemark[1],\newauthor
 Roberto Buonanno$^3$\footnotemark[1],  Gianni Marconi$^4$\footnotemark[1].\\ 
$^1$ Instituto de Astrof\'\i sica de Canarias. Calle V\'\i a L\'actea s/n. E38200 - La Laguna, Tenerife, Canary Islands, Spain.\\
$^2$ University of La Laguna. Avda. Astrof\'isico Fco. S\'anchez, s/n. E38206, La Laguna, Tenerife, Canary Islands, Spain.\\
$^3$ INAF--Osservatorio Astronomico di Teramo. Via Mentore Maggini, 64100 Teramo, Italy.\\
$^4$ ESO Vitacura. Calle Alonso de C\'ordova 3107, Vitacura, Casilla 19001, Santiago de Chile 19, Chile.}
\begin{document}

\date{Accepted 2013 May 8. Received 2013 May 7}

\pagerange{\pageref{firstpage}--\pageref{lastpage}} \pubyear{2013}

\maketitle

\label{firstpage}

\begin{abstract}
We present the Star Formation History (SFH) and the age-metallicity relation (AMR) in three fields of the Fornax dwarf spheroidal galaxy. They sample a region spanning from the centre of the galaxy to beyond one core radius, which allows studying galactocentric gradients. In all the cases, we found stars as old as 12 Gyr, together with intermediate-age and young stellar populations. The last star formation events, as young as 1 Gyr old, are mainly located in the central region, which may indicate that the gas reservoir in the outer parts of the galaxy would have been exhausted earlier than in the centre or removed by tidal interactions. The AMR is smoothly increasing in the three analyzed regions and similar to each other, indicating that no significant metallicity gradient is apparent within and around the core radius of Fornax. No significant traces of global UV-reionization or local SNe feedback are appreciated in the early SFH of Fornax.

Our study is based on FORS1@VLT photometry as deep as $I\sim$24.5 and the IAC-star/IAC-pop/MinnIAC suite of codes for the determination of the SFH in resolved stellar populations.

\end{abstract}


\section{Introduction}
\label{introduccion}

Galaxy formation and evolution is still an open question. In the $\Lambda$-CDM scenario, dwarf galaxies are the building blocks from which larger galaxies are formed \citep*{Blumenthal1985, Dekel1986, Navarro1995, Moore1998}. The dwarfs we observe today may be surviving systems that have not yet merged with larger galaxies and may contain the fossil record of the early evolution of galaxies. Evolution and star formation history (SFH) of these galaxies are very likely affected by local processes such as supernovae feedback and tidal interactions with nearby systems, as well as by global cosmic environmental factors like the early reionization of the Universe by UV radiation  \citep*{Taffoni2003, Hayashi2003, Kravtsov2004, Kazantzidis2011}. Characterizing their evolution and star formation history (SFH) could shed light on the physical mechanisms involved.\\

Dwarf galaxies are, by number, the dominant type of galaxies in the Local Group and provide an unique window onto galaxy formation. Their proximity allows us to resolve their individual stars, obtaining detailed measurements of the ages and chemical abundances of their populations. Dwarf Spheroidals (dSph) are systems characterized by low surface luminosities ($\rm \sum_V \la 0.002~L_\odot~pc^{-2}$), small sizes (about a few hundred of parsecs), and lack of gas. The relatively large velocity dispersions observed in the dSphs, exceeding 7 $\rm km~s^{-1}$ \citep*[see][and references therein]{Aaronson1983, Mateo1998}, suggest the presence of abundant dark matter in them. Assuming to be virialized systems, dSphs reach mass-to-light ratios ($M/L$) of $\sim 5 -500$ in solar units \citep{Kleyna2001, Odenkirchen2001, Kleyna2005}. All these discoveries have given rise to competing interpretations concerning their origin and cosmological significance.\\

In this work we present a study of the Fornax dSph, one of the nine classic dSph satellites of the Milky Way (MW). Located in the Fornax constellation, it was discovered by \citet{Shapley1938}. Several studies \citep*{Mackey2003, Greco2007, Tammann2008, Poretti2008, Greco2009} have used the RR-Lyrae or dwarf cepheids to calculate the distance to Fornax which ranges from 128 to 142 kpc. Its dynamical mass within the observed half light radius has been estimated in $\rm 5.6\times10^7 M_\odot$ \citep[][and references therein]{McConnachie2012}. Its principal barionic component is stellar, with an ambiguous detection of an $HI$ cloud of mass $\sim 0.14\times 10^6 M_\odot$ \citep*[][assuming a distance of 136 kpc]{Bouchard2006} offset from optical centre of the galaxy and possibly associated to the Milky Way.\\

\begin{table*}
\begin{minipage}{118mm}
\caption{Main data of The Fornax dSph.}
\label{tab:at_Glance}
\begin{tabular}{@{}lcr}
\hline
Quantity & Value & References\footnote{(1) \citet{van_den_Bergh1999};  (2) \citet{Mackey2003}; (3) \citet{Greco2007};  (4) \citet{Greco2009}; (5) \citet{Irwin1995}; (6) \citet{Mateo1998}; (7) \citet{McConnachie2012}; (8) \citet{Poretti2008}; (9) \citet{Tammann2008}; (10) \citet{Walker2006}.}\\
\hline
RA, $\alpha$ (J2000.0) &  2h 39$^\prime$ 53.1$^{\prime\prime}$ & (1)\\
Dec, $\delta$ (J2000.0) & -34$^{\circ}$ 30$^\prime$ 16.0$^{\prime\prime}$ & (1)\\
Galactic longitude, $l$ ($^\circ$) & 237.245 & \\
Galactic latitude, $b$ ($^\circ$) & -65.663 & \\
Heliocentric distance (kpc) & 136$\pm$5 & (2) (3) (4) (8) (9)\\
Heliocentric velocity (km s$^{-1}$) \hspace{20pt}& 55.3$\pm$0.1 & (7) \\
Ellipticity, $e$ & $0.30\pm 0.01$ & (5)\\
Position angle ($^\circ$) & 41$\pm$6 & (5) \\
Core radius ($^\prime$) & 13.8$\pm$0.8 & (6)\\
Tidal radius ($^\prime$) & 71$\pm$4 & (6)\\
Luminosity, $L_V$ ($L_\odot$) & $15.5\times 10^6$ & (10) \\
Barionic mass, $M_\star$ ($M_\odot$) & $2\times 10^7$ & (7) \\
Total mass at half light radius, $M_{\rm dyn}(\leq r_h)$ ($M_\odot$) & $5.6\times 10^7$ & (7) \\
\hline
\end{tabular}
\end{minipage}
\end{table*}

Fornax is, after the Sagittarius dSph, the largest and most luminous of the MW companions, with a core radius of $\sim$ 550 pc (using data listed in Table~\ref{tab:at_Glance}). These two galaxies are the only dSph satellites of the MW hosting globular clusters. Moreover, Fornax shows two conspicuous star clumps located at $17^\prime$ and at 1.3$^{\circ}$ from its centre. \citet{Coleman2004}, \citet{Coleman2005} and \citet{Coleman2008} proposed that these overdensities are shell structures resulting from a merger with a smaller, gas-rich system that occurred 2 Gyr ago. If this interpretation is correct, it may also explain the recent episodes of star formation. The merger scenario is also supported by  \citet*{Amorisco2012} who found  distinct radial velocity stellar components with different metallicities, and suggest that Fornax is a merger of a bound pair. The gas expelled by the first generations of stars would have been re-captured explaining the pre-enrichment observed in the inner clump \citep{Olszewski2006}. On another hand, \citet{deBoer2013} claims that these clumps are more likely to be the result of the quiet infall of gas previously expelled by Fornax during its star formation episodes. We summarize some of the main characteristics of the Fornax dSph in Table~\ref{tab:at_Glance}.\\

The SFH of the Fornax dSph galaxy has been obtained by \citet{Coleman2008} and more recently by \citet{deBoer2012}. In both papers several fields are analyzed, covering a large fraction of the galaxy body, extending up to about one tidal radius. In this paper we revisit the SFH, including the age-metallicity relation (AMR) of the central region of Fornax and up to a few arcmins beyond the core radius, based on deep FORS1@VLT photometry ($I \sim$24.5). The fact that our photometry is deeper than that of \citet{deBoer2012} allows us a more accurate and detailed analysis of the SFH, the AMR and their galactocentric dependences in the inner region of the galaxy. In $\S$\ref{Cap:Data} data set is presented. In $\S$\ref{Cap:CMD} the color magnitude diagrams (CMDs) for the observed regions are described. Section $\S$\ref{Cap:SFH_obtaining} describes the method followed to obtain the SFH. In $\S$\ref{Cap:SFH_results} the SFH is given. A comparison with previous works is given in $\S$\ref{Cap:Comparison}. In $\S$\ref{Cap:Discussion}, the results are discussed. Finally, a summary and the main conclusions of the work are presented in $\S$\ref{Cap:Summary}.\\

\section{The data}
\label{Cap:Data}

\subsection{Observations}

\begin{table*}
 \centering
 \begin{minipage}{165mm}
\caption{Observed fields and exposures.}
\label{tab:observations}
\begin{tabular}{@{}lccccccc}
\hline
\multirow{2}{*}{Field} & \multirow{2}{*}{No. f.\footnote{Number of science frames for each field.}} & \multirow{2}{*}{$\alpha$ (J2000)}& \multirow{2}{*}{$\delta$ (J2000)}& G.radius\footnote{Distance of each field centre to the centre of Fornax ($^{\circ}$)} & V--Images Exp. time & I--Images Exp. time & Seeing\footnote{Typical seeing for the set of images.}\\
 & & & & ($^\circ$) & (s) & (s) & ($^{\prime\prime}$) \\
\hline
Center field & 16 & 2h 40$^\prime$ 00$^{\prime\prime}$ & -34$^{\circ}$ 30$^\prime$ 17$^{\prime\prime}$ & 0.02 & 3700 (6$\times$600$+$2$\times$50) & 4600 (6$\times$750$+$2$\times$50) & 0.6\\
Outer field 1 & 11 & 2h 39$^\prime$ 40$^{\prime\prime}$ & -34$^{\circ}$ 20$^\prime$ 0$^{\prime\prime}$ & 0.18 & 1850 (3$\times$600$+$1$\times$50) & 4600 (6$\times$750$+$1$\times$50) & 0.5\\
Outer field 2 & 11 & 2h 40$^\prime$ 10$^{\prime\prime}$ & -34$^{\circ}$ 20$^\prime$ 0$^{\prime\prime}$ & 0.18 & 1850 (3$\times$600$+$1$\times$50) & 4600 (6$\times$750$+$1$\times$50) & 0.5\\
\hline
\end{tabular}
\end{minipage}
\end{table*}

\begin{figure}
\begin{center}
\includegraphics[ scale=0.75]{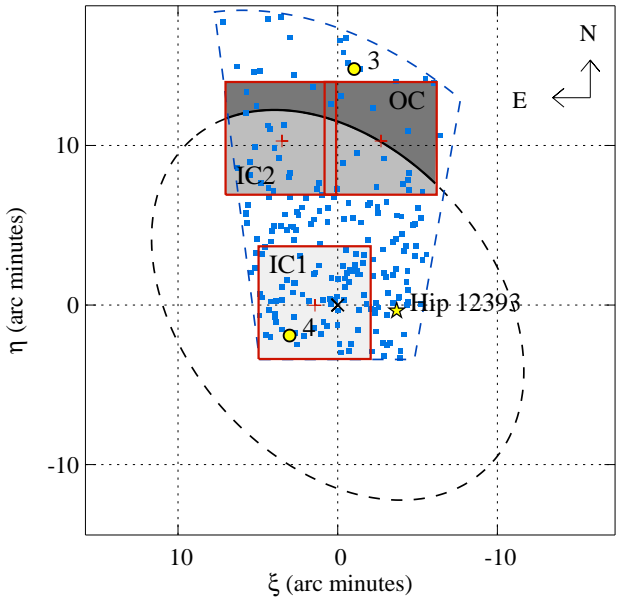}
\caption{Observed fields and the distribution of the CaT stars used for the spectroscopic AMR derivation in Fornax. Solid lines (red in the electronic version) show the boundaries of each observed field. Their centres are represented by crosses (red in the electronic version). The regions used for the study of SFH gradients are represented in different gray shades and labelled accordingly. CaT stars are represented by small squares (blue in the electronic version). These lie within the north section of an ellipse with semi-major axis of $20.7^\prime$ centered in Fornax, and the lower boundary of the IC1 region (blue-dashed line in the electronic version). The dashed ellipse corresponds to the core radius ($13.8^\prime$, position angle 41$^{\circ}$) given by \citet{Mateo1998}. The centre of Fornax is represented by a black cross. The star symbol shows the position of a bright foreground star (12393 Hipparcos). The yellow filled circles show the position of globular clusters Fornax 3 and Fornax 4 which 
are labelled accordingly.}
\label{fig:campos}
\end{center}
\end{figure}

The observations were performed in service mode in July 2000 with FORS1 at the \textit{VLT}. The \textit{FORS1} detector is a 2048 $\times$ 2048 pixel$^2$ CCD with a pixel size of $0.2^{\prime\prime}/pixel$, resulting in a total field of view of $6.8^\prime\times6.8^\prime$. A total area of 133.35 $(^\prime)^2$ was covered with three fields, one of them located at the centre of Fornax, the other two approximately located at $\sim10^\prime$ from the former; that is approximately one core radius, $r_{c}$, from the centre. The seeing was typically $0.6^{\prime\prime}$, but in some long exposures of 750 s, the seeing was as good as $0.4^{\prime\prime}$. Total integration times were 3700 s in $V$ and 4700 s in $I$ for the centre field. For the other two fields integration times were 1850 s in $V$ and 4600 in $I$ for each field. A schematic diagram of the observed fields is shown in Figure~\ref{fig:campos}. In Table~\ref{tab:observations} we show a summary of the observations.\\

\subsection{Photometry}
\label{Cap:Data,photo}

Data reduction was performed in the IRAF environment\footnote{IRAF is distributed by the National Optical Astronomy Observatories, which is operated by the Association of Universities for Research in Astronomy, Inc., under cooperative agreement with the National Science Foundation.}. Point-spread function (\textit{PSF}) fitting photometry was obtained using the \textit{DAOPHOT} / \textit{ALLFRAME} packages \citep{Stetson1987, Stetson1994}.\\

The photometric calibration was obtained comparing the magnitudes of our brightest and well measured stars with those of the calibrated photometric list by \citet{Stetson2000, Stetson2005}. For the calibration we used only stars with magnitudes within $16<I<21.25$ and a maximum photometric error of $\sigma=0.08$, both in our photometry and in the calibrated list. With this selection, stars close to saturation as well as low signal-to-noise stars are avoided. Since red stars exceed by far the blue ones, by number, we excluded some of the red-clump stars, keeping a reasonable homogeneity in the color distribution. This selection leaves us with 2795, 1216, and 1322 stars for calibration in the centre field and each of the outer fields respectively.\\

Both small gradients as a function of the pixel coordinates and color terms were found in the calibration. To correct these effects we performed two fits. First we applied a $3^{\rm th}$ degree spline fitting to the magnitude differences between our photometry and the calibrated one as a function of pixel coordinates. This seems a good compromise between accuracy and minimization of possible numerical artifacts. Second, color terms and photometric zero-points were obtained through a $2^{\rm nd}$ degree polynomial fitting to the corrected magnitudes obtained in the first step. Both fits were performed iteratively and applying $\sigma$-weighs and 3$\sigma$ clipping in each iteration.\\

\begin{figure}
\begin{center}
\includegraphics[scale=0.75]{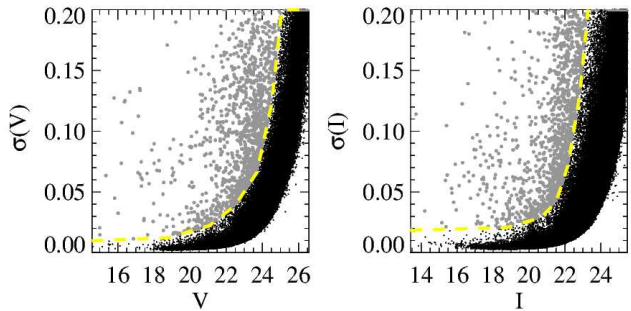}
\caption{$\sigma$ vs. magnitude of the measured stars. The dashed thick lines (yellow in the electronic version) show the $\sigma$ rejection limit adopted to clean the photometric list. Grey dots show the rejected stars. Similar selection was made using the \textit{SHARP} and $\chi^2$ parameters provided by \textit{ALLFRAME}.}
\label{fig:Errors_cut}
\end{center}
\end{figure}

The photometry was also corrected of distance modulus and reddening. We adopted a distance modulus $(m-M)_0=20.66 \pm 0.17$ (136$\pm $5 kpc), obtained as the weighted average value of the distances measured with RR-Lyrae or cepheids \citep{Mackey2003, Greco2007, Tammann2008, Poretti2008, Greco2009}. The reddening was corrected using dust maps by \citet*{Schlegel1998} for both filters ($V$,$I$). No substantial gradient is apparent in these maps within the boundaries of our observed fields. On another side, since the region of Fornax studied here lacks any significant amount of gas \citep{Bouchard2006} and the amount of interstellar dust is related to gas \citep*[see e.g.][]{Lisenfeld1998}, the internal reddening can be safely neglected. We therefore applied the mean reddening commonly assumed for Fornax, which is $E(V-I) = 0.028$ mag. Finally, we cleaned our photometry of likely non-stellar objects and stars with high magnitude errors on the basis of $\sigma$, $\chi^2$ and \textit{SHARP} parameters provided 
by \textit{ALLFRAME} for each star. Figure~\ref{fig:Errors_cut} shows $\sigma$ as a function of magnitude and the criteria used to select the final photometry list.\\

The \textit{observational effects} can impact the quality of the photometry of resolved stars. These include signal-to-noise limitations, detector defects, stellar crowding and all factors affecting and distorting the observational data with the resulting loss of stars, changes in measured colours and magnitudes, and systematic uncertainties. In order to quantify and characterize these observational effects we performed artificial stars tests on each frame. Details of the followed procedure are given by \citet{Hidalgo2011}. Suffice to mention here that, for the case of Fornax, a total of one million of artificial stars were injected in each field in successive steps of $\sim 3000$ artificial stars each, recovering their magnitudes with the same procedure as for the real stars. For each artificial star, we record the injected magnitude, $m_i$, the recovered magnitude, $m_r$, the position on the CCD, and a flag indicating whether the star was recovered, which will be used later on for the simulation of 
observational effects in synthetic CMDs.\\

\section{The CMDs of Fornax and the space distribution of stars}
\label{Cap:CMD}

The position of our observed fields allows us to study the SFH dependence on the distance to the galactic centre. We defined three different regions for our study. The core radius ellipse \citep{Mateo1998} was used to delimit them. Figure~\ref{fig:campos} shows the three selected regions. Details of them are given in Table~\ref{tab:regions}. It should be noted that the core radius ellipse is obtained from the spatial distribution of all the stars, which is different from that of young stars \citep*[e.g.,][]{Stetson1998}.\\

\begin{table*}
 \centering
 \begin{minipage}{120mm}
\caption{Geometric parameters of the three defined regions (See Figure~\ref{fig:campos}).}
\label{tab:regions}
	\begin{tabular}{@{}lcccc@{}}
	  \hline
	  \multirow{2}{*}{Name of Region} & Area\footnote{Area of each region after removing 273.2 pc$^2$ covered by globular custer Fornax 4} & Radial distance\footnote{Projected mean distance from the centre of Fornax to the baricentre of the stars in each region.} & \multirow{2}{*}{No. of stars\footnote{Total number of stars for each region.}} & \multirow{2}{*}{No. of RGB stars\footnote{RGB stars within $19.45\leq I \leq 21.7$. Their numbers are used to normalize the Hess CMDs (See Figure ~\ref{fig:CMDs}).}} \\
	  & (pc$^2$) & (pc) & & \\
	  \hline
	  Inside Core \#1 (IC1) & 75770.1  &   56.8  &  69590  &  370 \\
	  Inside Core \#2 (IC2) & 84248.6  &  360.4  &  69712  &  302 \\
	  Outside Core (OC)     & 59181.5  &  473.6  &  38643  &  117 \\
	  \hline
	  \textbf{Total:}       & 219200.2 &     -   &  177945 &  789 \\ \hline
	\end{tabular}
\end{minipage}
\end{table*}

\begin{figure*}
\begin{center}
\includegraphics[scale=0.75]{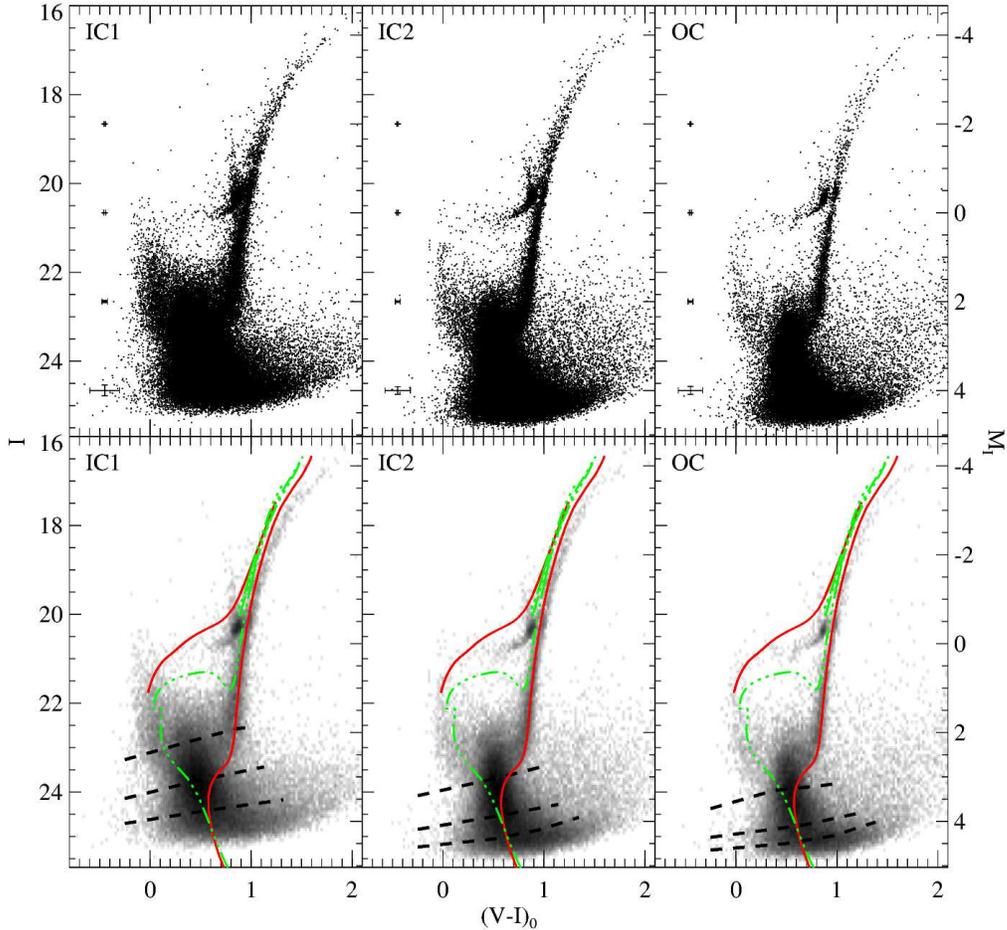}
\caption{The CMDs of the three regions (upper panels) and the corresponding Hess diagrams (bottom panels) in logarithmic gray scale. Apparent $I$ magnitudes are shown at the left vertical axis, while absolute magnitudes are shown in the right-side vertical axis. Average photometric $\sigma$ are shown for different $I$ in the left-side of each CMD panel (upper panels), while the dashed black lines mark, from top to bottom, the completeness levels of 90\%, 75\%, and 50\% of each CMD (bottom panels). Two isochrones from BaSTI stellar evolution library have been ploted over the Hess CMDs: Z=0.004, 1Gyr (dotted-dashed green line in the electronic version) and Z=0.001, 13.5 Gyr (solid-red line in the electronic version).}
\label{fig:CMDs}
\end{center}
\end{figure*}

The calibrated CMDs are shown in Figure~\ref{fig:CMDs}. Reaching a limmiting $I$ magnitude of $\sim 24.5$ $\sim 25.2$ and $\sim 25.3$ with a completeness of 50\% in the IC1, IC2 and OC regions respectively, these are by far, the deepest ground CMDs of Fornax. This allows to resolve the oldest main sequence turn-offs (oMSTO) with small errors and at the $\sim75$\% completeness level for the IC1 region, and $\sim90$\% for the outer regions.\\

Remarkable differences are found in the stellar population as moving outward. The most significant one regards the population of younger, blue main sequence (MS) stars, which are more abundant in the inner regions. Significant numbers of stars can be seen above the oldest turn-offs in all the CMDs, indicating an extended star formation in the three regions. However, less populated blue plume of the MS in the outer regions, suggests that the latest bursts of star formation were preferentially located toward the centre of Fornax. In addition, the red giant branch (RGB) is narrower in the outer regions, which may be related with a smaller metallicity dispersion (see Figure~\ref{fig:sfh3d-all} below). Small changes can also be marginally seen in the the horizontal branch (HB) and in the red-clump (RC) morphology in the sense that the HB might be relatively more populated in the outer regions.\\

\section{Obtaining the star formation history}
\label{Cap:SFH_obtaining}

\subsection{The method}

To solve the SFH in the three regions we followed the method described in \citet{Aparicio2009}. This method provides the Star Formation Rate (SFR) as a function of time and the AMR of the system. Three main codes are the mainstays of the method: (1) IAC-star \citep{Aparicio2004}, used for the computation of synthetic stellar populations and CMDs; (2) IAC-pop \citep{Aparicio2009}, for the computation of SFH solutions, and (3) MinnIAC \citep{Hidalgo2011}, a suite of routines created to manage the process of sampling, observational effects simulation, creating input data, averaging solutions and calculating uncertainties. Details of the method can be found in the aforementioned papers. A short summary is given here.\\

Considering that time and metallicity are the most important variables in the problem \citep[see][]{Aparicio2009}, we define the SFH as an explicit function of both. Formally, $\psi(t,Z){\rm d}t{\rm d}Z$ is the mass converted into stars within the time interval $[t,t+{\rm d}t]$ and within the metallicity interval $[Z,Z+{\rm d}Z]$. $\psi(t,Z)$ can be identified with the usual definition of the SFR, but as a function of time and metallicity.\\

A synthetic CMD (sCMD) is computed with IAC-star corresponding to a stellar population created with constant star formation rate and an uniform metallicity distribution within two fixed values over the full history of the galaxy (time from 0 to 13.5 Gyr). The limits of the metallicity distribution are chosen such that they encompass the full metallicity range expected for the analyzed system. This synthetic stellar population is divided according to selected age and metallicity bins into a set of simple stellar populations, $\psi_i$, each one formed by stars within small metallicity and time intervals. In this way, any SFH can be written as a linear combination with positive coefficients of the form $\psi(t,z)=\sum \alpha_i\psi_i$. Deriving the SFH of the analyzed object requires the determination of the $\alpha_i$ parameters. If we denote $M_i^j$ the distribution of stars in the CMD of the simple stellar population $\psi_i$, $M^j=\sum\alpha_i M_i^j$ will be the one corresponding to the CMD of the full SFH $\
psi(t,z)$. Here, the index $j$ account for the position (color and magnitude) on the CMD. The $\alpha_i$ parameters can be now determined comparing the stellar distribution in $M^j$ and the one in the observed CMD, $O^j$, using a merit function. The $\chi^2_\gamma$ defined by \citet{Mighell1999} is used.\\

The aforementioned comparison of CMDs requires a well performed sampling of the diagrams. It is, in particular, important to give a higher weight to stellar evolutionary phases providing more accurate information for the purpose of the SFH computation, to better sampled CMD regions, or to the ones with smaller observing errors. For this, we define several macro-regions in the CMD, \textit{bundles}, which delimit the evolutionary phases used for our analysis. At the same time, we divide each bundle into boxes using different samplings (See \citealp{Aparicio2009} and \citealp{Hidalgo2011} for details).\\

\subsection{The case of Fornax}

\begin{figure*}
\begin{center}
\includegraphics[scale=0.75]{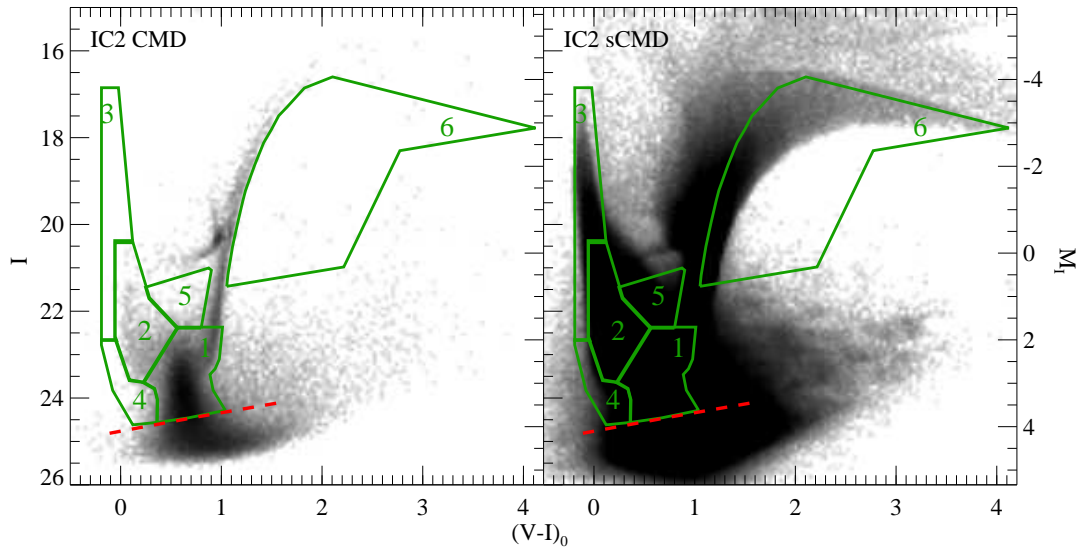}
\caption{Left: observed Hess CMD of the IC2 region. Right: Hess diagram of the sCMD computed for the IC2 region and including a simulation of the observational effects. In both cases a logarithmic grey scale has been used. The sampled regions (\textit{bundles}) are shown by green (in the electronic version) lines and labelled from 1 to 6. The completeness level of 75\% is shown by a red (in the electronic version), dashed line. Similar samplings were done for the CMDs of the IC1 and OC regions.}
\label{fig:bundles}
\end{center}
\end{figure*}

For the analysis of Fornax, we computed a sCMD with $3\times 10^7$ synthetic stars. We used the scaled-solar BaSTI \citep{Pietrinferni2004} stellar evolution library, completed by \citet{Cassisi2007} models for very low mass stars. For the metallicity range, we adopted a uniformly spread metallicity between $Z=0.0001$ and $Z=0.02$ (Solar metallicity). A fraction of 40\% of binaries was adopted \citep{Monelli2010a} and the IMF used was the one described by \citet{Kroupa2002}, i.e. a power law with exponent $x=2.3$ for stars with masses above $\rm 0.5~M_\odot$, and $x=1.3$ for stellar masses between 0.08--0.5 $\rm M_\odot$. For illustrative purposes, the sCMD obtained for region IC2 is shown in Figure~\ref{fig:bundles}.\\

Observational effects were simulated for each star in the sCMD using the information provided by the artificial stars tests described in $\S$\ref{Cap:Data,photo}. The simulations account for the differences between the injected ($V_i$, $I_i$) and recovered ($V_r$, $I_r$) magnitudes of the artificial stars as a function of magnitude, color, and position on the CCD, thus accounting for different crowding levels. For each synthetic star with magnitudes $V_s$ and $I_s$, a list of artificial stars is created with those fulfilling $|V_i - V_s|\leq\epsilon_V \wedge |I_i - I_s|\leq\epsilon_I \wedge d\leq \epsilon_d$, where $d$ is the spatial distance between the synthetic and the artificial star and $\epsilon_V$, $\epsilon_I$ and $\epsilon_d$ are free input searching intervals. Then, through a simple random sampling a single artificial star is selected from this list as individual representative for the synthetic star. If the star was unrecovered in some of the filters, the synthetic star is eliminated from the sCMD.
 Otherwise, considering $V_i^\prime$, $I_i^\prime$ and $V_r^\prime$, $I_r^\prime$ as the injected and recovered magnitudes of the selected artificial star, $V_s^e=V_s + V_i^\prime- V_r^\prime$ and $I_s^e=I_s + I_i^\prime- I_r^\prime$ will be the magnitudes of the synthetic star with observational effects simulated \citep[See][for details]{Hidalgo2011}.\\

To compare observed CMD and synthetic sCMD, we defined 6 bundles as shown in Figure~\ref{fig:bundles}. The sampling is as follow: (1) lower MS; (2) young MS; (3, 4) the boundary regions of the MS that may be expected to be occupied by the lowest metallicity stars; (5) the subgiant branch stars, and (6) the limit set by the higher metallicity stars of the RGB. In all cases, bundles are defined above the 75\% completeness level.\\

Sampling of the simple stellar population was undertaken using different binning depending on the considered age and metallicity ranges. For the age, several binnings of 0.25, 1 and 2 Gyrs wide were used between the age nodes 0, 0.5 11.5 and 13.5 Gyrs in the same order. For metallicity, the binnings were $2\times10^{-4}$, $5\times10^{-4}$, $1\times10^{-3}$ and $1\times10^{-2}$ wide within the nodes $1\times10^{-4}$, $3\times10^{-4}$, $5\times10^{-4}$, $1\times10^{-3}$, $1\times10^{-2}$ and $2\times10^{-2}$.\\

In order to minimize the dependence of the solution on the CMD sampling parameters, we obtained 36 solutions varying the CMD binning and the simple stellar population sampling. The average $\psi$ of the 36 solutions with its associated $\chi^2_\gamma$ (obtained as the average of the 36 single $\chi^2_\gamma$) is adopted as the solution. Besides this, in order to limit also the effects on the solution of uncertainties in photometric zero points, reddening, and distance to the galaxy, we computed the SFH, according to the procedure explained above, for 25 different offsets in color and magnitude applied to the three observed CMDs. Used offsets in color were $\Delta(m_v-m_i)=(-0.06, -0.03, 0, 0.03, 0.06)$, and $\Delta(I)=(-0.2, -0.1, 0, 0.1, 0.2)$ for the magnitude. In total, 900 solutions were computed for each region.\\

The shift with the best $\chi^2_\gamma$ found for region IC1 was different than the one obtained for regions IC2 and OC by about 0.03 mags in colour. The more accurate photometry of the external fields, less affected by crowding, leaded us to adopt their best shift ($\Delta(m_v-m_i)=0.04$ and $\Delta(I)=-0.033$) for the three regions, which is compatible within $\sim 1\sigma$ with the IC1 best $\chi^2_\gamma$ shift. This is also within the uncertainties of the photometric zero points, the measured distance modulus, and extinction toward Fornax. Furthermore, reddening maps by \citep{Schlafly2011} provides a slightly lower reddening value for Fornax ($E(V-I) = 0.025$) than the one adopted in the present work using \citet{Schlegel1998} maps, which would be in accordance with a color shift to the red.\\

Finally, for each region, the 36 solutions obtained for the adopted shift were finally boxcar smoothed in order to obtain the final SFH.\\

\begin{figure*}
\begin{center}
\includegraphics[scale=0.75]{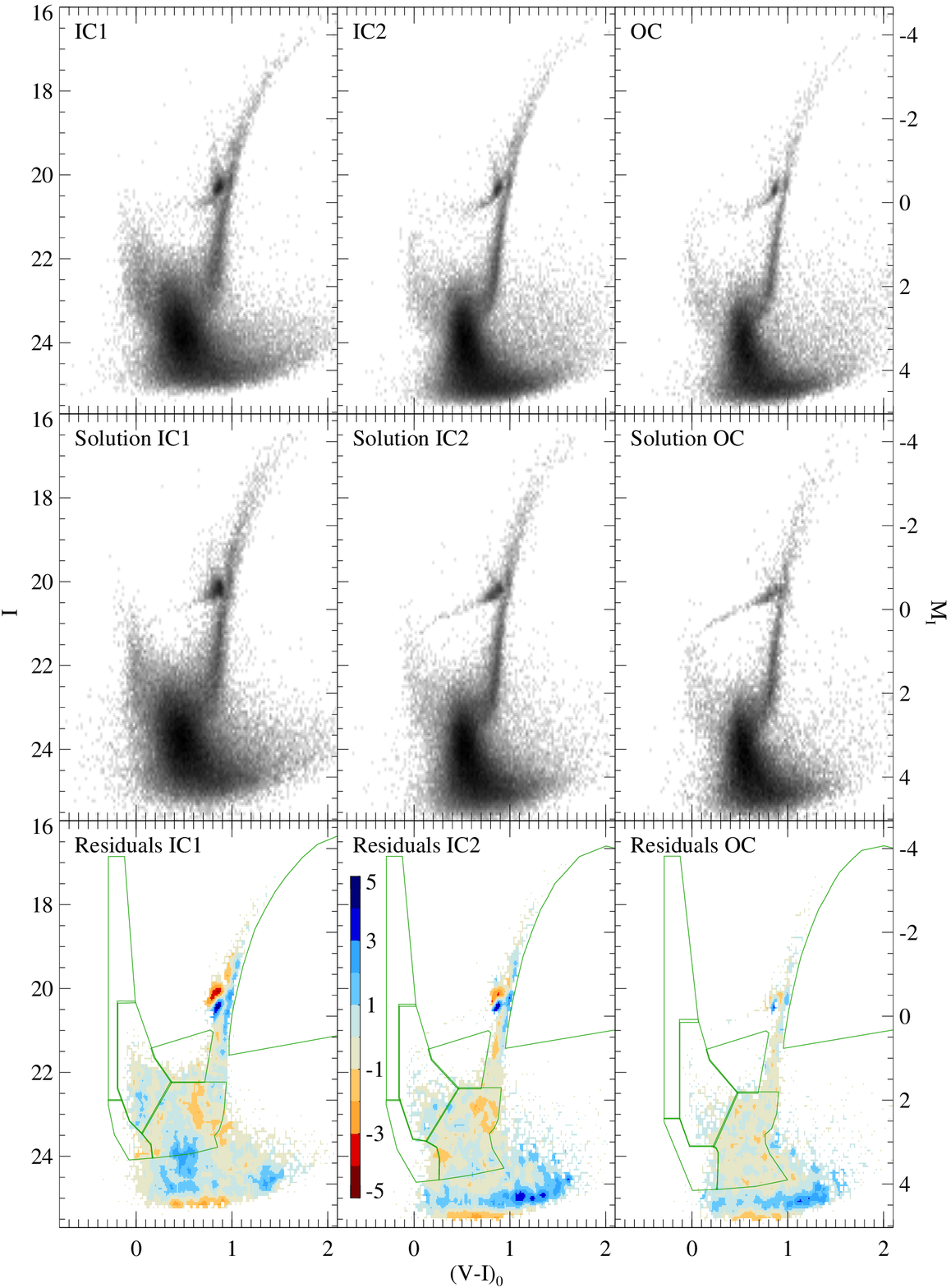}
\caption{Hess diagrams of the CMDs for observed stars (upper panels), best-fitting models (middle panels) and residuals (bottom panels). The gray scales are logarithmic for the observed and synthetic CMDs. The residuals are in units of Poisson uncertainties and represented in a red-blue scale (in the electronic version). Bundles sets are shown in green (in the electronic version).}
\label{fig:Obs-Sol}
\end{center}
\end{figure*}

Figure~\ref{fig:Obs-Sol} shows the Hess CMD corresponding to the solutions for each region and the residuals from their substraction to the corresponding observed CMD. Residuals are given in units of Poisson uncertainties, obtained as $(n_i^o-n_i^c)/\sqrt{n_i^c}$ where $n_i^o$ and $n_i^c$ are the number of stars in bin $i$ of an uniform grid defined on the observed and calculated CMDs, respectively. Only at the bottom of the MS and in the RC, both cases outside the used bundles sets, values outside $\pm 3 \sigma$ can be marginally apreciated. This indicate a good fitting of the model in all cases. These differences might be produced by the less precission of the stellar models during the helium core burning fase and by the lack of reliable information in the photometry below the 50\% of completeness level.\\

\section{The Star Formation History of Fornax: SFR and AMR}
\label{Cap:SFH_results}

\begin{figure}
\begin{center}
\includegraphics[scale=0.55]{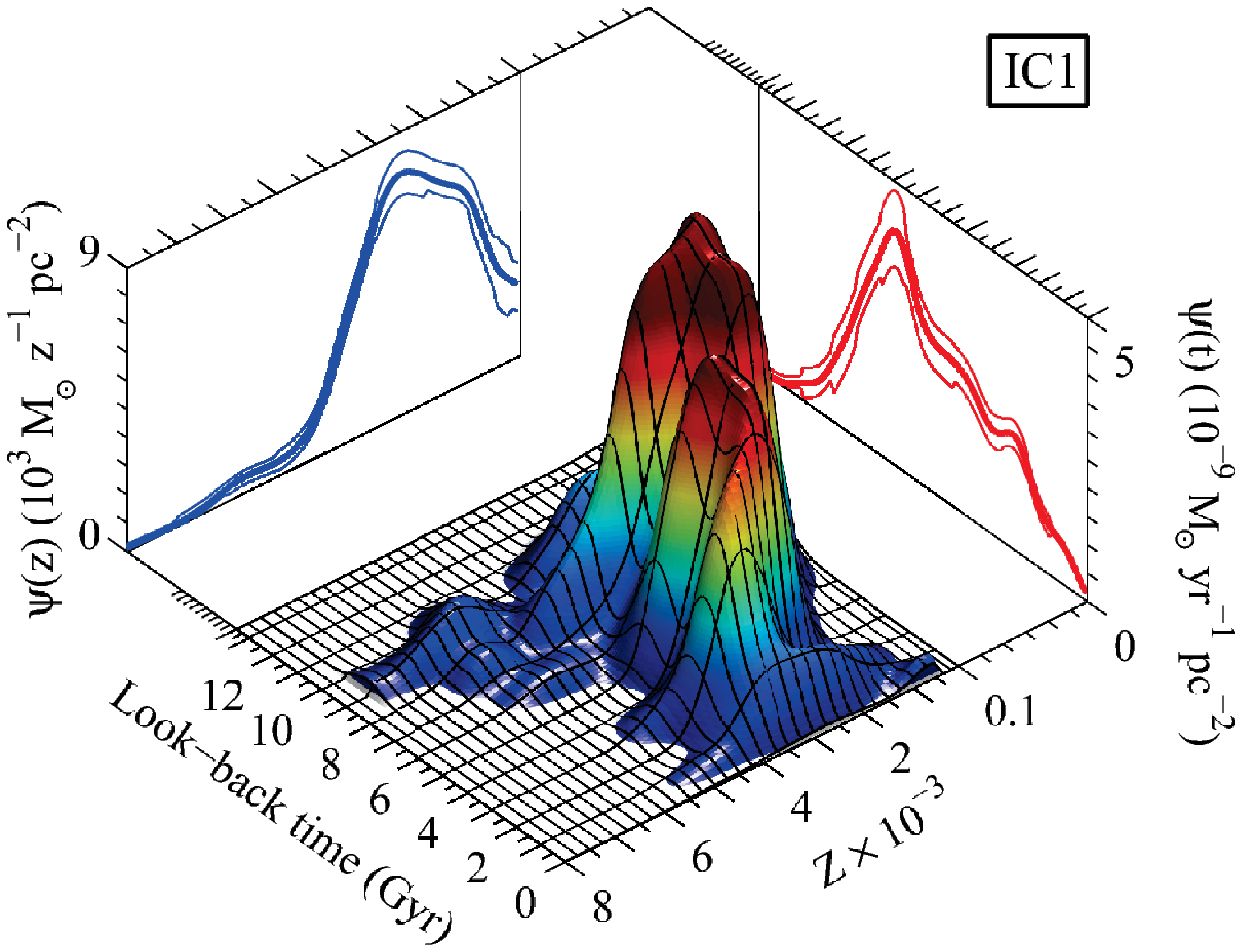}
\includegraphics[scale=0.55]{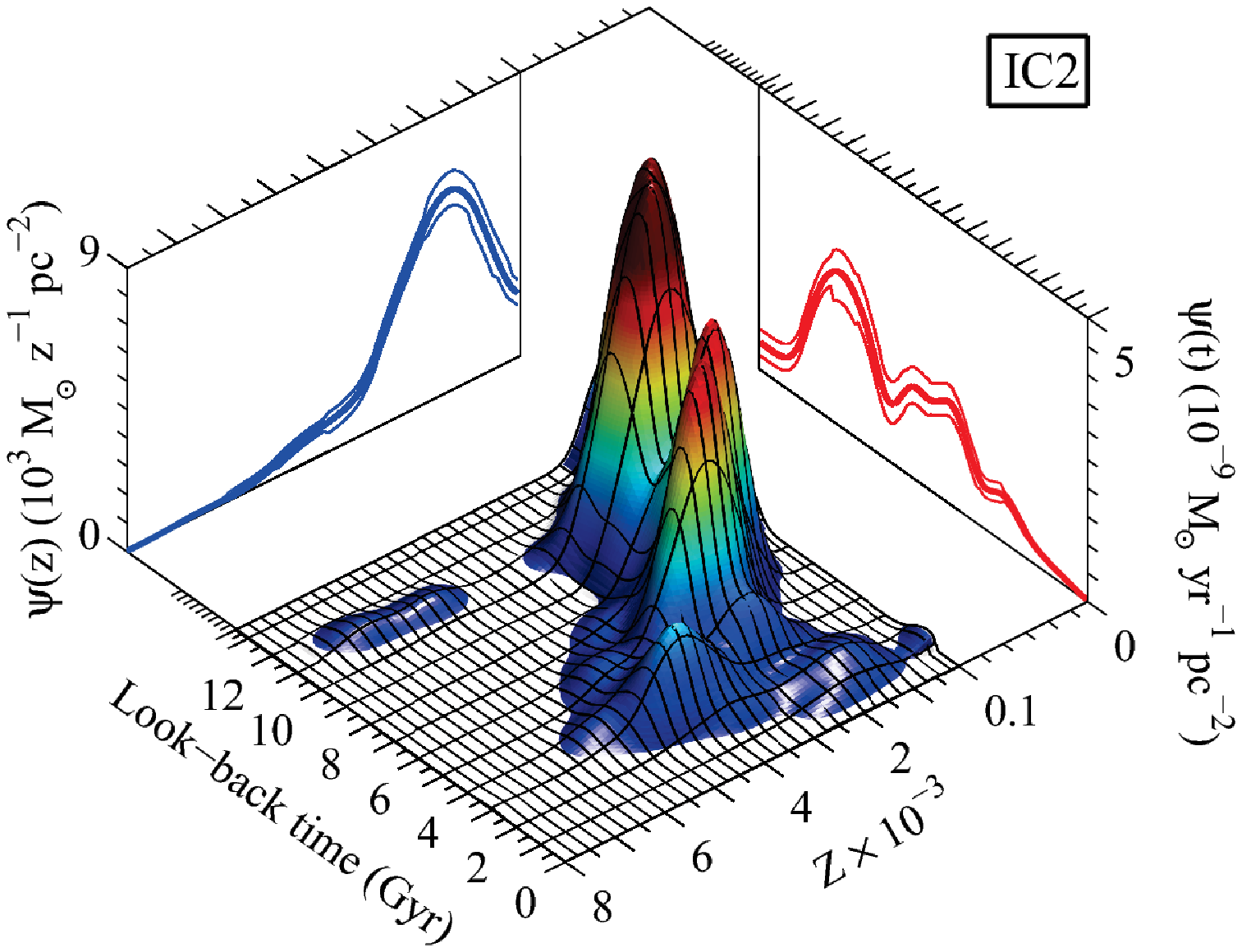}
\includegraphics[scale=0.55]{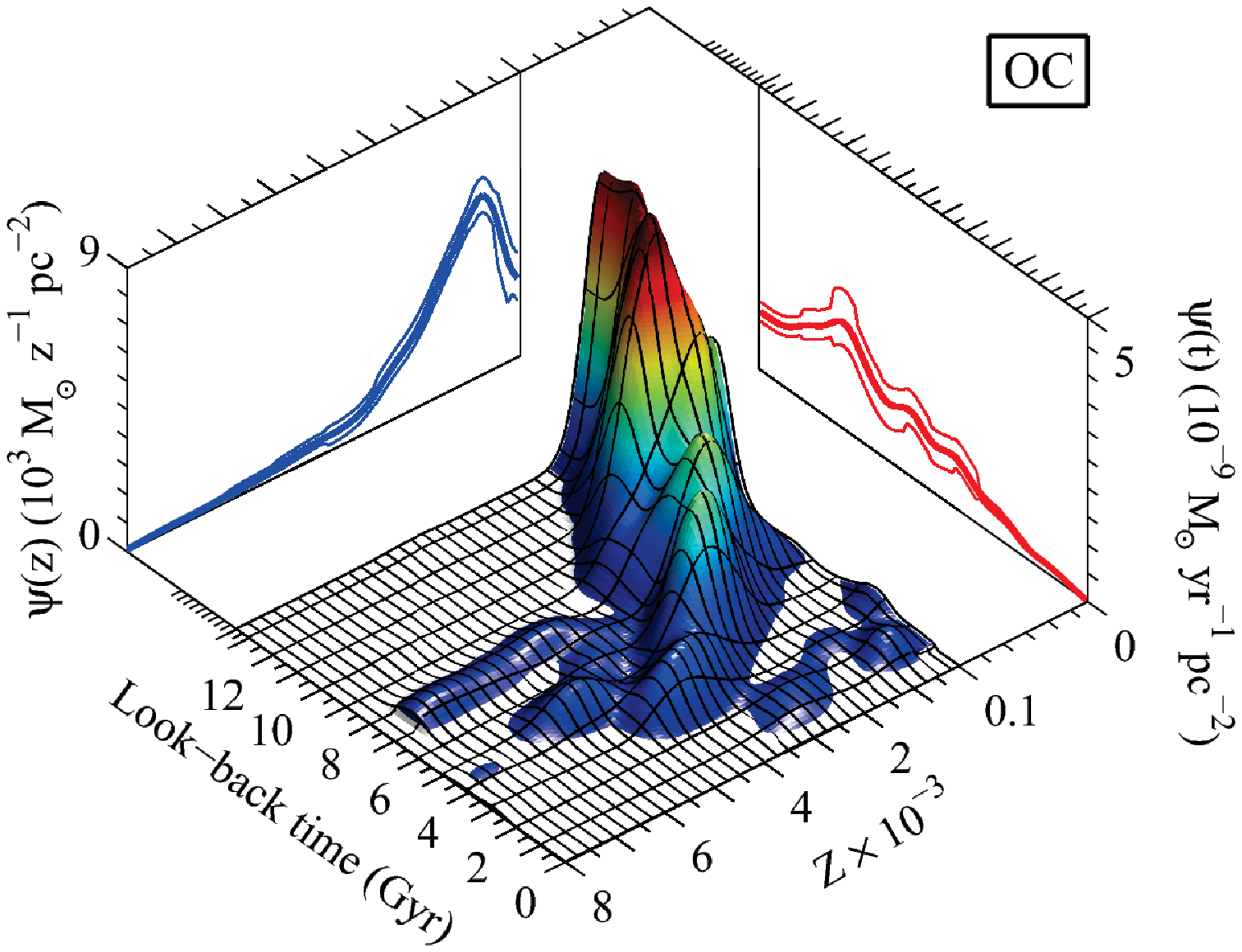}
\caption{The star formation as a function of time and metallicity for the three regions defined in Fornax. From top to bottom, the panels correspond to IC1, IC2, and OC. For each panel, the volume under the surface represents the mass converted into stars as a function of time and metallicity, $\psi(t,z)$. The left projection (blue curve in the electronic version) represents the stellar metallicity distribution at birth, while the projection on the right (red in the electronic version) shows the star formation rate as a function of time, $\psi(t)$. The age-metallicity relation can be obtained as a projection in the horizontal plane. Error intervals are represented by the dashed lines. Both, surfaces and curves for each region are normalized to the region area.}
\label{fig:sfh3d-all}
\end{center}
\end{figure}

\begin{figure}
\begin{center}
\includegraphics[scale=0.55]{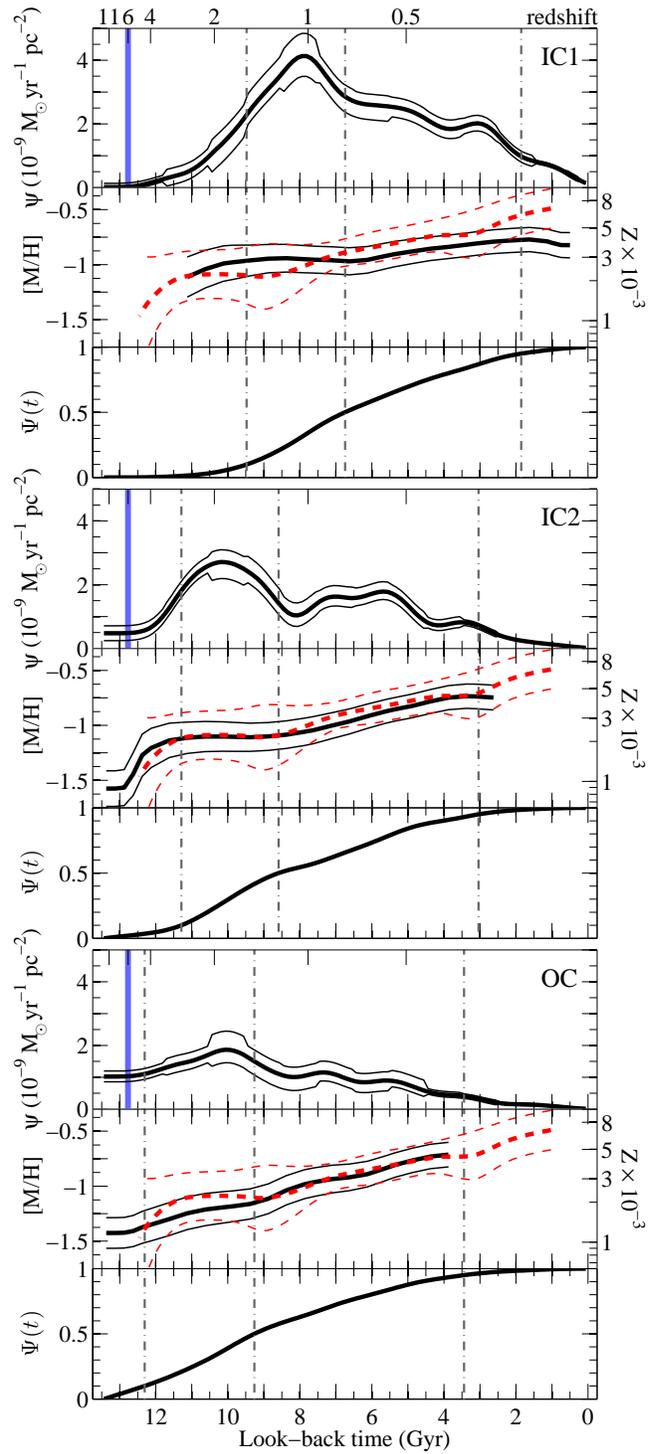}
\caption{Summary of results for the three analyzed regions. For each one, the star formation rate as a function of time ($\psi(t)$) (top), the age-metallicity relation (middle), and the cumulative mass fraction, $\Psi(t)$, (bottom) are shown. Error bands in $\psi(t)$ and dispersion in the AMR are drawn by thin lines. Units of $\psi(t)$ are normalized to the corresponding region area. 10th, 50th, and 95th percentiles of $\psi(t)$ are shown by dash-dotted vertical lines. A redshift scale is given in the upper axis, computed assuming $H_0=70.5$km $\times$ s$^{-1}$Mpc$^{-1}$, $\Omega_m$=0.273, and a flat universe with $\Omega_\Lambda$=$1-\Omega_m$. The end of UV-reionization epoch is marked at $z\sim6$ in blue (in the electronic version). The AMR obtained from CaT spectroscopy is also represented by a red-dashed line (in the electronic version).}
\label{fig:sfh2d-all}
\end{center}
\end{figure}

\begin{figure}
\begin{center}
\includegraphics[scale=0.55]{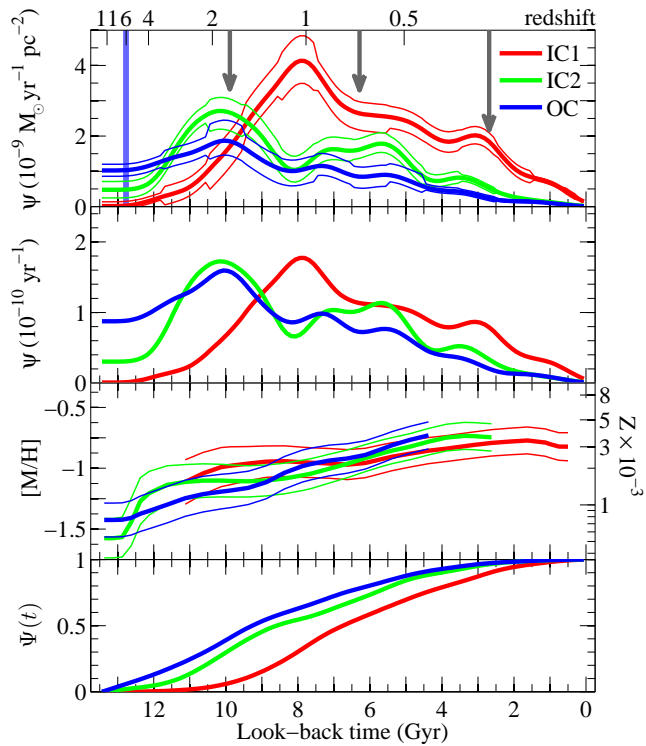}
\caption{Comparison between results shown in Figure~\ref{fig:sfh2d-all}: the star formation rate as a function of time ($\psi(t)$) (top panel); the same, but normalized to the temporal integral of $\psi(t)$ (second panel); the age-metallicity relation (third panel), and $\Psi(t)$, (bottom panel) for the studied regions IC1 (red in the electronic version), IC2 (green in the electronic version) and OC (blue in the electronic version) are shown. The end of UV-reionization era is marked at $z\sim6$ (blue in the electronic version). Possible passages of Fornax through it perigalacticon, derived from the orbital parameters by \citet{Piatek2007}, are represented by vertical arrows.}
\label{fig:sfh2d-superimposed}
\end{center}
\end{figure}

\begin{table*}
 \centering
 \begin{minipage}{105mm}
  \caption{Integrated Quantities derived for Fornax dSph.}
  \label{tab:Integrated_Quantities}
  \begin{tabular}{@{}lcccc}
    \hline
      \multirow{2}{*}{Region} & $\int{\psi(t')dt'}$ \footnote{Integrated between 0 and 13.5 Gyrs.} & $<\psi(t)>$  & $<age>$ & \multirow{2}{*}{$<Z>$} \\
      & $(10^6$ M$_\odot)$ & $(10^{-9}$ M$_\odot yr^{-1} pc^{-2})$ & $\rm(Gyrs)$ & \\
      \hline
      IC1 & $1.76 \pm 0.04$   &  $1.72 \pm 0.04$  & $6.4 \pm 0.7$ & $0.0025 \pm 0.0005$\\
      IC2 & $1.32 \pm 0.04$   &  $1.16 \pm 0.04$  & $8.1 \pm 0.8$ & $0.0020 \pm 0.0004$\\
      OC  & $0.69 \pm 0.02$   &  $0.87 \pm 0.03$  & $8.8 \pm 0.8$ & $0.0019 \pm 0.0005$\\
    \hline
  \end{tabular}
 \end{minipage}
\end{table*} 

Figure~\ref{fig:sfh3d-all} shows the resulting SFHs of the regions analyzed in the Fornax dSph as a function of both $t$ and $Z$, $\psi(t,Z)$. $\psi(t)$ and $\psi(Z)$ are also shown in the $\psi - t$ and $\psi - Z$ planes, respectively. The projection of $\psi(t,Z)$ on the age-metallicity plane provides the age-metallicity relation (AMR).\\

In Figure~\ref{fig:sfh2d-all} we summarize the main results for each region by showing $\psi(t)$ (top), the AMR (middle), and $\Psi(t)$ (bottom), defined as $\Psi(t)=\int_0^t \psi(t') dt'$ as a function of time. The AMR is obtained computing the average metallicity $Z$ as a function of time from the distribution function $\psi(t,Z)$; i.e. $<Z(t)>={\int Z'\psi(t,Z')dZ'}/{\int \psi(t,Z')dZ'}$. The metallicity dispersion as a function of time is also computed from $\psi(t,Z)$ as the interval around the average $Z$ containing 67\% of the distribution. In order to provide statistically reliable AMRs, we have only considered the time interval in which $\psi(t,Z) \geq 3\times 10^{-5}$ M$_\odot$yr$^{-1}$. In Figure~\ref{fig:sfh2d-superimposed} we show a complementary comparison of the results, including the estimated passages of Fornax through the perigalacticon of its orbit, as obtained from \citet{Piatek2007} orbital parameters (see below). In Table~\ref{tab:Integrated_Quantities} the mean integrated quantities 
derived for Fornax dSph are listed.\\

Together with Leo I \citep{Gallart1999}, Fornax shows the most recent star formation activity of any Local Group dSph, lasting near a Hubble time and holding stars about 1 Gyr old or even younger. The innermost region (IC1) has the youngest stellar population of the three regions, formed mainly by intermediate-age stars. In the outer regions (IC2, OC) the importance of recent star formation episodes decreases gathering the bulk of their stars at old-intermediate ages. While outer regions show a main burst occurred 10 Gyrs ago, the star formation in IC1 peaks $\sim$2 Gyrs later. In spite of having such different $\psi(t)$, all regions roughly show the same mean metallicity and a similar behavior in their AMRs. The metallicity starts to increase from the beginning of the SFH of Fornax, showing a smooth constant enrichment almost independently of the considered galactocentric distance. It is also noticeable that the oldest stellar populations in Fornax show already a significant metallicity enrichment. However can be, at least partially, been accounted for by the fact that the minimum metallicity in our models is $Z=0.0001$ together with the resolution of our solutions.\\

Regardless of the considered region we found a small amount of metal-poor but relatively young ($\sim$1--4 Gyrs old) stars in Figure~\ref{fig:sfh3d-all}. This population clearly does not follow the general AMR, and is likely to be blue straggler stars (BSs) population \citep{Momany2007, Mapelli2007, Mapelli2009, Monelli2012}. This is not in contradiction with the claim by \citet{Mapelli2009} that most BSs candidates in Fornax are young MS stars. Indeed, the BSs candidates defined by \citet{Mapelli2009} (see box 1 in their figure 2) approximately correspond to our bundle 2 (see Figure~\ref{fig:bundles}), which is contributing mainly to the young to intermediate-age population in our solutions.\\

We can extract additional information on the SFH gradients in Fornax from $\Psi(t)$ (lower panels of Figures~\ref{fig:sfh2d-all} and \ref{fig:sfh2d-superimposed}). In the case of the IC1 region, the fraction of total mass converted into stars increases at rather constant rate from 10 Gyrs ago to $\sim$2 Gyrs ago. As we move to outer regions, $\Psi(t)$ tends to rise earlier, reaching higher fractions of mass in shorter times. This can be shown by the time  corresponding to $\Psi=0.50$. It is $\sim$6.7 Gyrs ago in the IC1 region, $\sim$8.6 Gyrs ago in IC2 and $\sim$9.3 Gyrs ago in the OC region. Times corresponding to $\Psi=0.10$ and $\Psi=0.95$ can be used as estimates of the times by when the star formation starts and finishes in each region of Fornax. $\Psi=0.10$ occurred 9.50, 11.25, and 12.30 Gyrs ago and $\Psi=0.95$ corresponds to 1.85, 3, and 3.5 Gyrs ago in IC1, IC2 and OC respectively. All these values agree on indicating that star formation stopped earlier in the outer regions.\\

\subsection{Estimating time resolution in the SFH}

Observational uncertainties as well as the uncertainties inherent to the SFH computation procedure result in a smoothing of the derived SFH and a decrease of the age resolution. Having an accurate estimate of the latter is fundamental to ascertain the reliability of the solutions and to better characterize the properties of the galaxy. We have calculated the age resolution as a function of time by recovering the SFH of a set of mock stellar populations corresponding to very narrow bursts ($10^6$ years) occurring at different times. Four mock populations have been computed, each one with a single burst at ages of 3, 8, 10, and 12.77 Gyr (the end of UV-reionization epoch, see below). We used IAC-star to create the CMDs associated with each mock stellar population. For each mock population, fifteen simulations have been done using different random number seeds and total stellar masses ranging from $\sim 1\times10^6$ to $0.1\times10^6$ M$_\odot$ in order to better characterize them. Observational 
uncertainties 
were simulated in the CMDs by using the results of the DAOPHOT completeness tests for each region. The SFHs of the mock stellar populations were then obtained using identical procedure as for the real data. In all cases, the recovered SFH is well fitted by a Gaussian profile with $\sigma$ depending on the burst age and metallicity. From these fits we estimate an intrinsic age resolution of $\sim 1.7$ Gyr at the oldest ages which improves to $\sim 0.6$ Gyr at an age of 3 Gyr. In all cases, the recovered ages are within $\pm 0.3$ Gyrs of the input ages, except for the 10 Gyrs burst in the IC1 region, which was recovered with a shift of $\sim 0.7$ Gyrs toward older ages. For the 3 Gyrs burst, shifts are lower than $\sim 0.1$ Gyrs in all the cases.\\

These results lead us to conclude that the main burst in the central region of Fornax was indeed delayed. The first important event of star formation occurred $\sim$10 Gyrs ago in the outer regions. This was followed by the main burst occurred in the IC1 region $\sim$7.5--8 Gyrs ago. There are very old stars in Fornax notwithstanding. In the three studied regions a small population of very old stars ($\geq$ 12 Gyrs) is detected, but the proportion of these decrease when moving to the inner parts of the core. This is consistentn with the presence of HB stars in the CMD of all regions.\\

\section{The SFR and AMR of Fornax: comparison with previous works}\label{Cap:Comparison}

There is a large database of metallicity measurements in Fornax stars from low and medium resolution spectroscopy of the infrared Ca$\rm_{II}$ triplet lines (CaT) \citep{Pont2004, Battaglia2006,Battaglia2008,Kirby2010,Letarte2010,Starkenburg2010, deBoer2012}. Results from these works are consistent to each other in general. Specifically, \citet{Kirby2010} offer a comparison of their metallicities obtained through medium resolution spectroscopy with those obtained from high resolution spectra \citep[][among others]{Letarte2010} in their Figure 16, while \citet{Letarte2010} compare their results with those from \citet{Battaglia2006} in their Figures 9 and 17. In turn, the work by \citet{deBoer2012} is based on the Kirby et al.'s, Letarte et al.'s and Battaglia et al.'s catalogues, and provide another comparison in their Figure 2. The only exception to the general good agreement is in \citet{Pont2004}, whose high metallicity range ($[Fe/H] > -0.75$) calibration appears to be not compatible with the rest, 
showing a significantly large number of stars with metallicities above $[Fe/H] = -0.5$.\\

A complete analysis of those results is beyond the scope of this work, still is very illustrative a comparison between the AMR obtained here, and other obtained using spectroscopic data. Since the works mentioned above cover a large field, a new set of stars have been selected for the comparison with our results. The stars have been collected from the \citet{Kirby2010}, \citet{Battaglia2006} (with the CaT $[Fe/H]$ calibration from \citet{Starkenburg2010}), and \citet{Letarte2010} data sets. From each catalogue, we selected the stars within the elliptical section shown in Figure \ref{fig:campos}. The total number of selected stars is 207, keeping a good statistical significance and avoiding spatial biases.\\

\citet{Kirby2010}, \citet{Battaglia2006} and \citet{Letarte2010} do not provide information about individual stars ages. For this comparison the stellar ages have been derived using the relationships by \citet{Carrera2008a,Carrera2008b}. Basically, the age of each star is derived from its metallicity and position in the CMD, both $V$ vs. $(V-I)$ and $V$ vs $(B-I$), through polynomial relationships. These relationships were obtained from a synthetic CMD created with IAC-star adopting the BaSTI stellar evolution library. The age determination uncertainties were obtained from a Monte Carlo realization stochastically varying the input parameters $[Fe/H]$, $\rm M_V$, $(V-I)$ or $(B-I)$ according to Gaussian probability distributions with sigmas similar to the observational uncertainties of these quantities. We refer the reader to \citet{Carrera2008a,Carrera2008b} for a detailed discussion of the procedure used, uncertainties, and test of confidence levels. In Figure~\ref{fig:sfh2d-all} the AMR obtained from the 
CaT sample is plotted. The spectroscopy based AMR and the ones obtained from our method agree within the errors in all cases. Only for the IC1 region, different trends may exist for the younger end of the AMR. The lower statistical significance of our AMR in this age interval and the possible contribution of BSs stars may account for the apparent disagreement.\\

Our results can also be compared with those derived by \citet{Coleman2008} and \citet{deBoer2012}. In particular the most relevant figures for our purposes are Figure 8 of \citet{Coleman2008} and Figure 16 of \citet{deBoer2012}. To do such comparison it is useful to take into account that our regions roughly extend from 0 to 0.1$^\circ$ (IC1); from 0.13$^\circ$ to 0.23$^\circ$ (IC2), and from 0.23 $^\circ$ to 0.28$^\circ$ (OC). In other words, our region IC1 is fully within the region 1 of \citet{Coleman2008} and within {\it annul1} region of \citet{deBoer2012}; our IC2 region is still within the region 1 of \citet{Coleman2008} but between the {\it annul1} and {\it annul2} of \citet{deBoer2012}, and our OC region is fully within the region 2 and region {\it annul2} of the respective papers. Results of \citet{Coleman2008} and \citet{deBoer2012} are consistent to each other. In particular they show a main burst occurred 3--4 Gyrs ago, and a metallicity increase from -1.25 at 12.5 Gyrs ago to -0.4 at present 
days. While our AMR (Figure~\ref{fig:sfh2d-all}) is very similar to those of these two papers, this is not the case for the SFH. Their results point to a younger stellar population, with the SFR (our $\psi(t)$) peaking at about 3.5 Gyrs ago for regions 1 and 2 in the case of \citet{Coleman2008} and at 4 and 5 Gyrs ago for the inner and the second annulus of \citet{deBoer2012}, with a SFR very low for ages older than 10 Gyrs in all the cases. In contrast, our $\psi(t)$ maxima occur at about 8 Gyrs ago for the IC1 and about 10 Gyrs for regions IC2 and OC and, although lower, the SFR is important for ages older than 10 Gyrs (see Figure~\ref{fig:sfh2d-all}). These differences are significant, and we think that they can be produced by the shallower photometry of \citet{Coleman2008} and \citet{deBoer2012}. On another side, \citet{Coleman2008}, and \citet{deBoer2012} find a clear stellar age galactocentric gradient, with the SFR peaking at older ages for larger galactocentric distances, in qualitative agreement 
with what we find here for the central regions.\\

Finally, using HST data, \citet{Buonanno1999} analyzed the surrounding field of the globular custer Fornax 4, which is included within our IC1 region. Fitting isochrones, they found stars with ages between 12 and 0.5 Gyrs, born mostly in several main bursts occurred roughly 7, 4 and 2.5 Gyrs ago, in good agreement with our results.\\ 

\section{Local and cosmic effects in the SFH of Fornax dSph galaxy}\label{Cap:Discussion}

\subsection{Local effects}\label{Cap:Local}

It is arguable that interactions with other systems would have affected the SFH of Fornax. These interactions would include possible mergers with other systems and tidal interactions with the MW. We believe that nothing clear in favor or against the relevance of these interactions or mergers on the evolution of Fornax can be concluded from our results. However, we think it is illustrative to make a short outline of the properties of Fornax that could be related with those interactions.\\

Firstly, the apparent lack of gas \citep{Bouchard2006} and important recent star formation in the outer regions of Fornax, could be due to gas removing by tidal interactions with the MW. But it may also be the result of quiet depletion of the gas reservoir in the outskirts of galaxy. Tidal interactions with the MW may also have triggered intense star formation events in Fornax that could be now detected in its SFH. To research this possibility, we used the orbital elements derived by \citet{Piatek2007} for Fornax in order to roughly estimate its perigalacticon passages. Assuming a stable orbit since 12 Gyr ago and adopting a heliocentric distance for Fornax of 136 kpc and 8.5 kpc for the distance of the Sun to the Galactic centre, the obtained values are $\sim$2.6, $\sim$6.2, and $\sim$9.8 Gyrs ago. These times are represented by vertical arrows in Figure~\ref{fig:sfh2d-superimposed}. A relation between $\psi(t)$ enhancements and perigalacticon passages is not clear, although it can not be ruled out from 
these data.\\

Secondly, it is remarkable that despite more than 10 Gyr of significant star formation activity, the AMR of Fornax is quite flat and the average metallicity has remained low. This behavior may be consistent with a merger process. If Fornax have suffered mergers with more metal-poor systems, the accretion of this fresher gas would have quenched the chemical enrichment. The observed delay in the main burst of the central region with respect to the surrounding regions, could be also the consequence of a major merger occurred at $z\sim1$. During this event the fresher fallen gas would have stayed confined within the central region of Fornax, enhancing its star formation and flattening the AMR as can be appreciated in Figure \ref{fig:sfh2d-all} for the IC1 region. However, the results obtained here are also qualitatively compatible with a scenario in which mergers are not relevant. On the one side, the observed delay in the main burst of the central region and the higher SFR, may be the natural result of a quiet, 
secular outside-in evolution. On another side, slow chemical enrichment could be produced by the galaxy being unable to efficiently retain the enriched material ejected from the stars.\\

\subsection{Global effects}\label{Cap:Global}

Besides interactions with other systems, two main processes can dramatically affect the formation and early evolution of dwarf galaxies: heating from the ultraviolet (UV) radiation arising from cosmic reionization and internal SN feedback. In principle, these processes can suppress the star formation in a dwarf galaxy and even remove its gas.\\

Evidence from quasar spectra indicates that the Universe was fully reionized by $z \sim $ 6, corresponding to a look-back time of $\sim$12.7 Gyr \citep{Becker2001}. Observing the cumulative mass fraction from Figure~\ref{fig:sfh2d-superimposed} it is clear that most stars in the Fornax regions studied in this paper were formed after reionization. In particular, $\sim$100\% of the stellar mass in IC1, $\sim$95\% in IC2 and  $\sim$90\% in OC were formed after $z=6$.\\

SN feedback does not appear to have a determining effect on Fornax either. The star formation smoothly decline after the main star formation peak, becoming flat in some regions. If the mechanical energy produced by SNe had significantly blown away the gas in Fornax, star formation would have strongly declined or even stopped after the main peak, as it can be observed in Cetus or Tucana dwarf galaxies \citep{Monelli2010a,Monelli2010b}.\\

\citet{Sawala2010} computed a number of galaxy evolution models for different total masses. They computed several subsets, one of them for the case that only SN feedback is at work, another with SNe feedback and UV radiation and a third one with the former effects plus self-shielding, which is probably the most realistic scenario. In the latter case, a total mass of at least $\sim 8\times 10^8$ M$_\odot$ is required in order to galaxies keep forming stars at a significant rate after UV-reionization. This is consistent with \citet{Walker2006} results, who claims that the total mass of Fornax could be as large as $10^9$ M$_\odot$.\\

\citet{Mayer2010} models indicate that dSph galaxies of the mass of Fornax would have had an initial mass significantly larger than the current one, and that their SFH would be shaped mainly by the combined effect of low star formation efficiency and gas stripping by ram pressure produced along pericenter passages, rather than by the combination of SNe feedback and UV background. In this sense, galaxies in the mass range of Fornax would have an extended star formation if their pericenter distance is relatively large or if their first pericenter passage is delayed to after $z=1$. The results obtained here for Fornax are consistent with this scenario.\\

\section{Summary and Conclusions}
\label{Cap:Summary}

The SFH has beed obtained for three regions located in the central part of the dSph galaxy of Fornax. The IAC method developed by \citet{Aparicio2009} to derive the SFH has been used.\\

Our results indicate that Fornax is a complex system, showing the most recent star formation events and the most extended SFH of any Local Group dSph. We find stars as young as $\leq$1 Gyr in Fornax, while the first star formation event occurred at the oldest star formation epoch ($\geq$ 12 Gyrs ago). The most recent star formation events are mainly located in the innermost region of the galaxy, which shows an important intermediate-age to young stellar population, while the region close to but beyond the core radius contains only old to intermediate-age stars. This may indicate that gas reservoir in the outer parts of the galaxy would be exhausted earlier than in the centre or removed by tidal interactions.\\

We have used two methods to obtain the AMR. On the one side, besides the SFR, the IAC method provides the AMR. Furthermore, we have obtained it from CaT spectroscopy of stars within the core radius, using data by \citet{Battaglia2006}. Both methods provide consistent results, within the error bars. The AMRs obtained for the three regions are smoothly increasing and similar to each other, indicating that no significant metallicity gradient is apparent within and around the core of Fornax.\\

The SFH of Fornax is little affected by SNe feedback and cosmic UV-reionization. This would be consistent with models by \citet{Sawala2010} in the case that the total mass of Fornax would be above $8\times 10^8$ M$_\odot$, as suggested \citet{Walker2006}. On another hand, accordingly to models by \citet{Mayer2010}, Fornax should have had its first perigalacticon passage after $z=1$ in order to have such extended star formation history.\\

\section*{Acknowledgments}

Support for this work was provided by the IAC (grant 310394), the Education and Science Ministry of Spain (grants AYA2007-3E3506 and AYA2010-16717). The data presented in this paper were obtained as part of a joint project between the University of Chile and Yale University funded by the Fundaci\'on Andes. The authors thank the anonymous referee for the comments that have helped to improve the paper.\\

\label{lastpage}
\end{document}